\documentclass[conference, onecolumn]{IEEEtran}
\usepackage{cite}
\usepackage{subcaption,graphicx}
\usepackage{amsmath}
\usepackage{algorithm}
\usepackage{algorithmic}

\makeatletter
\newcommand\fs@norules{\def\@fs@cfont{\bfseries}\let\@fs@capt\floatc@ruled
  \def\@fs@pre{}%
  \def\@fs@post{}%
  \def\@fs@mid{\kern3pt}%
  \let\@fs@iftopcapt\iftrue}
\makeatother
\restylefloat{algorithm}
\usepackage[shortlabels]{enumitem}

\IEEEoverridecommandlockouts
\begin{document}
%
% paper title
% Titles are generally capitalized except for words such as a, an, and, as,
% at, but, by, for, in, nor, of, on, or, the, to and up, which are usually
% not capitalized unless they are the first or last word of the title.
% Linebreaks \\ can be used within to get better formatting as desired.
% Do not put math or special symbols in the title.
\title{Energy-Efficient User Clustering for UAV-Enabled Wireless Networks Using EM Algorithm 
}

% author names and affiliations
% use a multiple column layout for up to three different
% affiliations

\author{

\IEEEauthorblockN{Salim~Janji}
\hspace{-1.5cm}\IEEEauthorblockA{\textit{\small Poznan University of Technology}\\
\small Poznan, Poland \\
\small salim.janji@doctorate.put.poznan.pl}
\and

\IEEEauthorblockN{Adrian~Kliks}
\hspace{-1.5cm}\IEEEauthorblockA{\textit{\small Poznan University of Technology}\\
\small Poznan, Poland \\
\small adrian.kliks@put.poznan.pl}
\thanks{Copyright © 2021 IEEE. Personal use is permitted. For any other
purposes, permission must be obtained from the IEEE by emailing pubs-
permissions@ieee.org. This is the author’s version of an article that has
been published in the 2021st International Conference on Software, Telecommunications and Computer Networks (SoftCOM) and published by IEEE. Changes were made to this version by the publisher prior to publication, the final version of record is
available at: http://dx.doi.org/10.23919/SoftCOM52868.2021.9559068.
To cite the paper use: S. Janji and A. Kliks, "Energy-Efficient User Clustering for UAV-Enabled Wireless Networks Using EM Algorithm," \textit{2021 International Conference on Software, Telecommunications and Computer Networks (SoftCOM)}, 2021, pp. 1-6, doi: 10.23919/SoftCOM52868.2021.9559068. or visit
https://ieeexplore.ieee.org/document/9559068}
}

% conference papers do not typically use \thanks and this command
% is locked out in conference mode. If really needed, such as for
% the acknowledgment of grants, issue a \IEEEoverridecommandlockouts
% after \documentclass

% for over three affiliations, or if they all won't fit within the width
% of the page, use this alternative format:
% 
%\author{\IEEEauthorblockN{Michael Shell\IEEEauthorrefmark{1},
%Homer Simpson\IEEEauthorrefmark{2},
%James Kirk\IEEEauthorrefmark{3}, 
%Montgomery Scott\IEEEauthorrefmark{3} and
%Eldon Tyrell\IEEEauthorrefmark{4}}
%\IEEEauthorblockA{\IEEEauthorrefmark{1}School of Electrical and Computer Engineering\\
%Georgia Institute of Technology,
%Atlanta, Georgia 30332--0250\\ Email: see http://www.michaelshell.org/contact.html}
%\IEEEauthorblockA{\IEEEauthorrefmark{2}Twentieth Century Fox, Springfield, USA\\
%Email: homer@thesimpsons.com}
%\IEEEauthorblockA{\IEEEauthorrefmark{3}Starfleet Academy, San Francisco, California 96678-2391\\
%Telephone: (800) 555--1212, Fax: (888) 555--1212}
%\IEEEauthorblockA{\IEEEauthorrefmark{4}Tyrell Inc., 123 Replicant Street, Los Angeles, California 90210--4321}}

% use for special paper notices
%\IEEEspecialpapernotice{(Invited Paper)}

% make the title area
\maketitle

% As a general rule, do not put math, special symbols or citations
% in the abstract
\begin{abstract}
Unmanned Aerial Vehicles (UAVs) can be used to provide wireless connectivity to support the existing infrastructure in hot-spots or replace it in cases of destruction. UAV-enabled wireless provides several advantages in network performance due to drone small cells  (DSCs) mobility despite the limited onboard energy. However, the problem of resource allocation has added complexity. In this paper, we propose an energy-efficient user clustering mechanism based on Gaussian mixture models (GMM) using a modified Expected-Maximization (EM) algorithm. The algorithm is intended to provide the initial user clustering and drone deployment upon which additional mechanisms can be employed to further enhance the system performance. The proposed algorithm improves the energy efficiency of the system by 25\% and link reliability by 18.3\% compared to other baseline methods.  
\end{abstract}

% no keywords

% For peer review papers, you can put extra information on the cover
% page as needed:
% \ifCLASSOPTIONpeerreview
% \begin{center} \bfseries EDICS Category: 3-BBND \end{center}
% \fi
%
% For peerreview papers, this IEEEtran command inserts a page break and
% creates the second title. It will be ignored for other modes.
\IEEEpeerreviewmaketitle

\section{Introduction}
% no \IEEEPARstart
Drone small cells (DSCs) have gained popularity in recent years as a solution to wireless communication problems such as lack of fixed infrastructure (e.g. due to natural disasters) or the need for temporal capacity increase (e.g. to manage traffic in a mass-event). If properly deployed, and despite its inherent limitations, UAV-enabled wireless can leverage the performance of the network due to increased probability for line-of-sight (LOS) connectivity and reduced total path loss. Furthermore, due to their mobility, the location of the BS can adapt to the changes in ground user distribution to improve the total throughput of the system. DSCs can also be a cost-effective substitute for building expensive cellular towers and infrastructure where the need for such infrastructure is limited in terms of time or capacity \cite{Mozaffari2019}, or be deployed along with existing infrastructure in heterogeneous cellular networks \cite{Chakareski2019}.

Nonetheless, utilizing UAVs for serving ground users in a given area has its limitations. Limited onboard energy restricts the deployment duration and transmit power thereby limiting the communication range. Furthermore, in the case of multiple DSCs, severe co-channel interference can substantially degrade users' link qualities\cite{abeywickrama}.  Therefore, the problem of 3D deployment and user allocation is a complex one when taking into consideration factors such as power minimization and interference mitigation. DSCs locations should minimize the path loss and maximize signal-to-interference-plus-noise-ratio (SINR) along with serving the highest possible number of users. Additionally, DSCs should be aware of each other to avoid inter-UAV collisions \cite{Mozaffari2019}.
\par The topic of DSCs deployment optimization gained a lot of focus in recent years [4-10]. The problem is commonly approached by optimizing a subset of all inherent aspects (e.g. locations or number of drones vs. coverage, transmit power vs. interference, etc.) without taking into account other factors such as initial deployment and user clustering. In this paper, we target the problem of power minimization, interference mitigation, DSC localization, and user clustering jointly along with deciding a minimum required number of drones to serve a given user set with known locations. Our centralized algorithm is based on Gaussian Mixture Models (GMM) and we call it Drone Users Clustering Expectation Maximization (DUCEM) algorithm. We believe it could also serve as an initial clustering step before applying further heuristic mechanisms to further improve system performance. That is, its results are more optimal than other clustering algorithms (e.g. K-means).  
\par The paper continues as follows. Section \ref{Related Work} briefly summarizes related research found in the literature. The system model and optimization problem formulation are given in Section \ref{System Model}. We introduce EM algorithm in \ref{EM} and present our modified version for DSCs clustering in Section \ref{DUCEM}. Section \ref{Simulation} introduces random waypoint mobility (RWM) model and reference point mobility group (RPGM) model that are used for user mobility simulation. We also describe K-means algorithm which is chosen as the baseline with which we compare our algorithm's performance and present the results. Finally, in Section \ref{conclusion}, we give our final remarks and conclusion. 

\section{Related Work} \label{Related Work}
The application of DSC has been investigated so far in various contexts. In particular, the authors in \cite{Hu2019} developed a machine learning scheme for predicting the number of required drones for load balancing in small cells based on historical data. In \cite{Lyu2015}, DSCs sequential placement is optimized to reduce the total number of drones and achieve maximum coverage. However, interference was not taken into account. The authors in \cite{Mozaffari2015} considered optimal deployment of two DSCs taking into consideration coverage and interference. In \cite{AffinityPropagation}, the authors proposed an interference mitigation scheme using affinity propagation which reduces the power of interfering drones and clusters the users using K-means algorithm without targeting the initial user clustering and allocation problem. In \cite{Kumbhar2018}, the authors mitigate DSC interference in LTE hetNets by applying 3GPP Release-10 enhanced inter-cell interference coordination (eICIC), and optimize UAV deployment using a genetic algorithm (GA). In \cite{Chakareski2019}, the authors optimize the transmit power and altitude of each DSC operating in a heterogeneous LTE network comprised of macro base stations, mmWave small base stations, and DSCs operating in the microwave band. The proposed algorithm determines subcarrier allocation using the Hungarian optimization method without targeting the problem of user-cell association. Using game theory, the authors in \cite{Li2020} formulated the problem in a mean-field game framework where the altitude of DSCs is controlled to improve total SINR. In \cite{Shi}, a solution based on Particle Swarm Optimization (PSO) algorithm was derived that maximizes coverage while maintaining link quality. 

In general, the mechanisms reported in the literature do not jointly consider the following dynamic issues of DSCs deployment: 
\begin{itemize}
    \item number of required drones for a given area
    \item interference mitigation between DSCs
    \item limited energy of DSCs.
\end{itemize}
Our work presents a heuristic mechanism that tries to address all of the aforementioned aspects of DSCs deployment.

% \begin{figure}[htp]
%   \subfloat{
%     \includegraphics[clip,width=\columnwidth]{dsigmaNone.jpg}}
%     \subfloat{
%     \includegraphics[clip,width=\columnwidth]{dsgima.jpg}}
%  \caption{\textcolor{red}{Comparison of clustering overlap for a given set of test points with different $d\Sigma_{max}$ values.}}
% \end{figure}

\section{Problem Formulation} \label{System Model}
Fig. 1 presents the considered scheme where $U$ ground users are randomly distributed and served by the set of $M$ drones, $ \mathcal{M}$. We consider a 3D Cartesian coordinates system in which each user $u$  is located at $\boldsymbol x_u= (x_u, y_u)$.  The location of the $j$th drone is denoted by $\boldsymbol F_j=(x_j,y_j,h_j)$, where $h_j$ denotes the flying attitude, and $j \in  \mathcal{M}$. In our system, we consider that all active drones are flying at the same fixed height $h = 10 \;m$. As discussed in \cite{10m}, $10 \;m$ is the optimal height for positioning a typical small cell antenna. Lower values cause possible coverage issues, and higher ones increase interference with neighbouring cells. Next, we assume  drone $j\in \{1,..,M\}$ can transmit at a specified power $P_{T_j}\leq P_{\max}$, and $P_{\max}$ is the maximum transmit power of the DSC. In consequence, it will serve all $U_j \leq U$ users within its coverage area. As will be explained later in Section \ref{DUCEM}, $P_{T_j}$ is directly proportional to $\Sigma_j$, the covariance value of the Gaussian distribution associated with drone $j$.

\begin{figure}[htbp]
\centering
\includegraphics[width=3in]{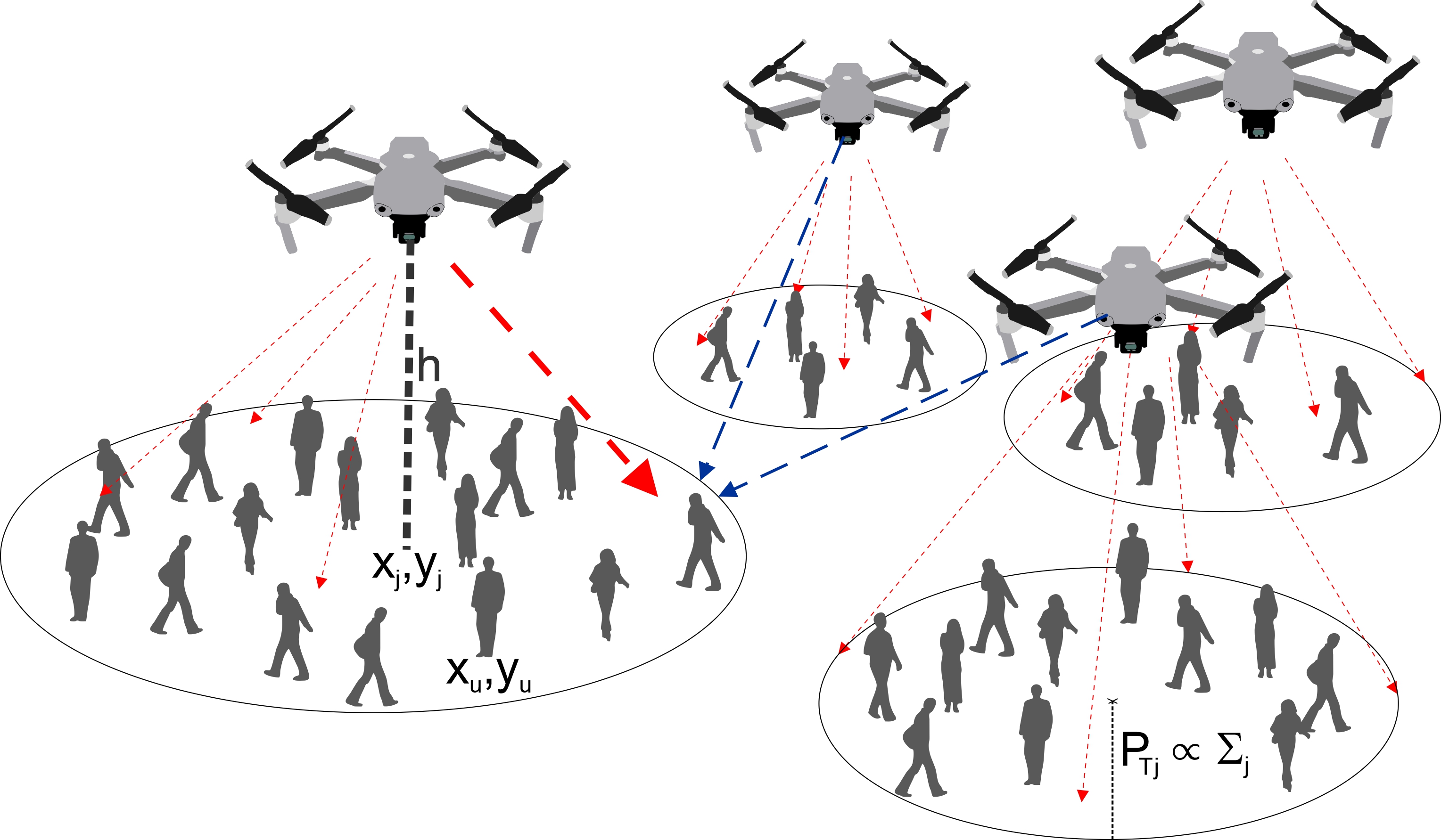}
\DeclareGraphicsExtensions.
\caption{Scenario of users served by DSCs.}
\label{fig_sim}
\end{figure}
As the drones are flying over a certain area, it is rational to assume that a LOS link exists between drones and users \cite{Wu2018}. Thus, we evaluate the free-space path loss model in our analysis. The distance between each drone and ground user is given by:
\begin{equation}
    d_{j,u}=\sqrt{(x_j-x_u)^2+(y_j-y_u)^2+h^2},\label{eqn:dist}
\end{equation}

\noindent and the channel gain from drone $m$ to user $u$ can be defined as:
\begin{equation}
    H_{j,u}=\alpha_0d^{-\lambda}_{j,u}.\label{eqn:plos}
\end{equation}

\noindent Here, $\lambda$ is the path loss exponent and $\alpha_0$ is the channel power at reference distance $d_0=1\;\mathrm{m}$. In consequence, the signal-to-noise ratio (SNR) between user $u$ and serving drone $j$ is then given by:
\begin{equation}
    \mathrm{SNR}_{j,u}= \frac{P_{T_j} H_{j,u}}{\sigma_0^2}
\end{equation}
\noindent where $\sigma_0^2$ is the power of additive white Gaussian noise (AWGN) at the receiver. 
%\subsection{Link Quality}
%Supposing user $u\in U_m$ such that the set $U_j$ denotes the users served by drone $m$. 
For modeling the interference from other DSCs, the signal-to-interference-plus-noise ratio (SINR) of user $u$ served by DSC $j$, $\Gamma_{u,j}$, is given by:
\begin{equation}
    \Gamma_{u,j} = \frac{P_{T_j} H_{j,u}}{\sigma_0^2+\sum_{i\neq j}^M P_{T_i} H_{i,u}}.
\end{equation}
In consequence, the theoretical capacity limit can be derived based on Shannon's formula as 
%\subsection{Total Throughput}
%The total throughput of the system is given by:
\begin{align}
    R_{\mathrm{total}}&=\sum^{M}_{j=1}\sum^U_{u=1} W(j,u)B \;\log_2(1+\Gamma_{u,j}),
\end{align}
where $W(j,u) = 1$ if user $u$ is served by drone $j$ and $0$ otherwise, and $B$ is the allocated bandwidth per user channel.

%\subsection{Power Consumption}
Focusing only on data communications, the total maximum power consumed by drones can be given by:
\begin{equation}
    P_{total}=\sum^M_{i=1}P_{T_i},
\end{equation}
where we have neglected propulsion energy as it is irrelevant to our problem.
%\subsection{Energy Efficiency}
Thus, the energy efficiency of the whole system is defined as follows:
\begin{align}
    EE&=\frac{R_{\mathrm{total}}}{P_{\mathrm{total}}}.
\end{align}

%\subsection{Energy Optimization Problem}
It is assumed that the transmitted power of each  drone can be continuously varied up to a limit $P_{\max}$. Also, the maximum number of users served by each drone is also limited to $U_{\max}$. Furthermore, each user is guaranteed a minimum $\mathrm{SNR}$ value of ${\mathrm{SNR}}_T$. The optimization problem is then formulated as:
\begin{align}
    \max_{P_{T_j}, F_j, U_j} &EE = \max_{P_{T_j}, F_j, U_j} \frac{R_{\mathrm{total}}}{P_{\mathrm{total}}}.
\end{align}
$s.t.$ 
\begin{enumerate}[a)]
    \item $\mathrm{SNR}_{j,u}\geq{\mathrm{SNR}}_T \;\;\; \forall u\in \mathcal{U} $
    \item $ U_j \;\;\; \leq U_{\max} \;\;\; \forall j\in  \mathcal{M}$
    \item $P_{T_j} \leq P_{\max} \;\;\; \forall j\in \mathcal{M} $.
\end{enumerate}
Next Sections present our heuristic approach to solving this problem.

\section{Expected-Maximization Algorithm} \label{EM}
Expectation Maximization (EM) algorithm is based on Gaussian Mixture Models (GMMs) where it is assumed that the set of samples is drawn from a "mixture" of Gaussian distributions \cite{MIT}. A GMM can be defined as:
\begin{equation}
    \pi(\boldsymbol x;\theta)=\sum^M_{j=1} P(j)N(\boldsymbol x;\boldsymbol\mu_j,\boldsymbol\Sigma_j),
\end{equation}
where the parameters $\theta = \{P, \boldsymbol\mu, \boldsymbol\Sigma \}$ include mixing proportions,  means of Gaussian components, and covariances respectively, and $N(\boldsymbol x;\boldsymbol\mu_j,\boldsymbol\Sigma_j)$ represents the $j$-th Gaussian distribution. \par
The EM algorithm learns the parameters $\theta$ given a set of sample points, $\boldsymbol X_t$, and number of mixtures, $M$ (As will be explained later, M denotes the number of drones). It consists of two steps (estimation step, called E-step, and maximization step, called M-step) performed iteratively: 
\begin{enumerate}
    \item (E-step) Evaluate the posterior assignment probabilities given by:
    \begin{align}
        p^{(l)}(j|t) &= p(j|\boldsymbol x_t,\theta^{(l)})=\nonumber\\
        &=\frac{P^{(l)}(j)N\left( \boldsymbol x_t;\boldsymbol \mu_j^{(l)},\boldsymbol \Sigma_j^{(l)}\right)}{\sum^M_{j^\prime=1}P^{(l)}(j^\prime)N\left(\boldsymbol x_t;\boldsymbol \mu_{j^\prime}^{(l)},\boldsymbol \Sigma_{j^\prime}^{(l)}\right)}\nonumber \\
        &=\frac{P^{(l)}(j)N\left(\boldsymbol x_t;\boldsymbol \mu_j^{(l)},\boldsymbol \Sigma_j^{(l)}\right)}{P(\boldsymbol x_t;\theta^{(l)})}
    \end{align}
    where $(l)$ denotes iteration number, and $\boldsymbol x_t = \left[ {\begin{array}{cc}
   x_t  \\
   y_t \\
  \end{array} } \right]$ is the sample $t = \{1,..,n\}$ where $n$ is the number of samples.
    \item (M-step) Update parameters according to:
    \begin{align}
        p^{(l+1)}(j)&= \frac{\hat{n}(j)}{n}, \;\; \text{where } \hat{n}(j)= \sum^n_{t=1}p^{(l)}(j|t).\\
        \boldsymbol \mu_j^{(l+1)}&= \frac{1}{\hat{n}(j)}\sum^n_{t=1}p^{(l)}(j|t)\boldsymbol x_t.\\
        \boldsymbol \Sigma_j^{(l+1)}&=\\
        \frac{1}{\hat{n}}\sum^n_{t=1}&p^{(l)}(j|t)
        \left(\boldsymbol x_t-\boldsymbol \mu_j^{(l+1)}\right)\left(\boldsymbol x_t-\boldsymbol \mu_j^{(l+1)}\right)^T.
        \nonumber
    \end{align}
\end{enumerate}
As can be observed, the
E-step obtains the posterior probabilities that denote the probability of having the sample observed $x_t$ being drawn from the $j$-th Gaussian component. In the M-step, the parameters of each component $j$ are updated to increase the total likelihood of the GMM. It is worth noting that the EM algorithm is guaranteed to converge to a local optimal solution \cite{MIT}.

\section{Drone Users Clustering Expected Maximization Algorithm (DUCEM)} \label{DUCEM}

In this paper, we propose using a modified EM algorithm to tackle the problem of DSCs placement and user clustering along with considering the system energy efficiency. 
\par The set of sample points, $\boldsymbol X$, is taken to be the set of users distributed on the ground to be served by DSCs. Furthermore, $\boldsymbol \mu_j$ indicates the $j$-th DSC location in the 2D $x,y$ Cartesian plane, and $\boldsymbol \Sigma_j$ is associated with the transmission power $P_{T_j}$ of drone $j$ (i.e. the values of each covariance matrix are proportional to the transmission power of each DSC). The posterior probabilities $p(j|t)$ bear no mapping to parameters in the real system but serve in determining the power and location of each DSC as evident in equations (12) and (13).
\par The reasoning behind our choice is that by examining equations (10-13) we observe that each sample point $\boldsymbol x_t$ (i.e. location of user $t$) contributes to the parameters of all components (drones). The magnitude of this contribution is determined by the posterior probability given in equation (10). Thus, the contribution of $\boldsymbol x_t$ to DSC $j$ parameters decreases as $p(k|t)_{k\neq j}$ increases. This is evident from (10) as $ N(\boldsymbol x;\boldsymbol \mu_k,\boldsymbol \Sigma_k)$ increases, $ p(j|t)$ decreases. This translates into the observation that user $t$, if served by DSC $k$, would contribute less to parameters updates of DSC $j$ in (11-13). Therefore, because of EM algorithm dynamics, it can be associated with the practical problem of DSC users clustering where each user requires to be served by one DSC. 
Our DUCEM algorithm introduces several modifications to the standard EM to suit our considered problem:
\begin{enumerate}
    \item Each covariance matrix $\boldsymbol \Sigma_j{_{M\times M}}$, is a diagonal matrix with equal diagonal values to map the DSCs circular coverage of ground users to a circular distribution of samples around means (power is transmitted equally through each axis around the DSC). By removing this constraint, coverage of DSCs will take a non-circular form (i.e. using beamforming). This will be a topic of our future work.
    \item The diagonal values of $\boldsymbol \Sigma_j$, denoted as $\Sigma_j$,  are limited to a maximum value $\Sigma_{\max}$ to resemble the real case scenario of limited transmit power of drones $P_{\max}$. This limits the number of sample points $\boldsymbol x_t$ to be associated with component $j$.
    \item  $\Sigma_{j}$ is only allowed to increase while $U_j \leq U_{\max}$. This ensures that the number of users served by DSC $j$ is bounded by $U_{\max}$.
    \item The change of $\Sigma_{j}$ per DUCEM iteration is limited by a maximum value of $d\Sigma_{\max}$. This reduces overlapping between clusters and therefore the chance for a cluster to be completely contained in another one. Not limiting $d\Sigma_{\max}$ allows abrupt big jumps in $\Sigma_{j}$ initial updates where sample points far from $\boldsymbol \mu_{j}$ substantially affect the update in (13) because the corresponding posterior probability is large (other components have not yet contributed to the denominator in (10)). Fig. 2 shows the difference between setting a maximum $d\Sigma_{max}$ for a certain set of sample points and $M=4$. As observed, limiting $\Sigma_j$ change in equation (13) reduces overall overlapping. Clearly, this improves interference between DSCs. To summarize, the update in equation (13) is limited by $\boldsymbol \Sigma_j^{(l+1)} - \boldsymbol \Sigma_j^{(l)} \leq d\Sigma_{max}$, and $\Sigma_j^{(l+1)} \leq \Sigma_{\max} $.
    \item Each user is associated to the cluster with smallest value of Standarized Euclidean Distance (SED) given by:
    \begin{equation}
        S\!E\!D_{j,u}=\frac{\sqrt{((x_j-x_u)^2+(y_j-y_u)^2+h^2)}}{\Sigma_j}.
    \end{equation}
    Users allocations are given by the matrix $\boldsymbol W_{M\times U}$ where $\boldsymbol W(j,u) = 1$ if the $u$-th user is associated with drone $j$ and $0$ otherwise.
    \item Each DSC is centered at the mean location of its served users.
    \item The algorithm begins with $m=1$ as the initial number of drones (components) and adds another drone while a solution that satisfies the requirements of the maximum number of users or link reliability is not yet found. 
    \item At each iteration, the total system EE is obtained. The setting that yields maximum EE ($\theta_{opt} = \{ \boldsymbol \mu_1,  \Sigma_1 ,...,\boldsymbol \mu_M,  \Sigma_M\}$ is saved and is chosen as the final solution. EM algorithm converges when the maximum change,  $maxd\theta$, of any of the parameters $\boldsymbol \mu_j$ and $  \Sigma_j$ after iteration $i$ is limited by $\varepsilon_1$. Also, a new DSC is added when the number of DSCs is not enough to find a solution that satisfies the maximum number of users and maximum power constraints. Finally, $\varepsilon_2$ ensures that no new drones are added until EM updates are close to convergence. 
\end{enumerate}
\begin{figure}[htbp]
\centering
\subfloat[High overlap for $d\Sigma_{max}=\infty$]{%
\centering
  \includegraphics[clip]{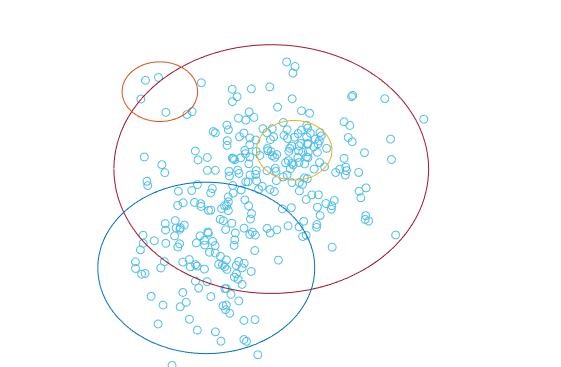}%
}

\subfloat[Decreased overlap for $d\Sigma_{max}=0.1$]{%
\centering
  \includegraphics[clip]{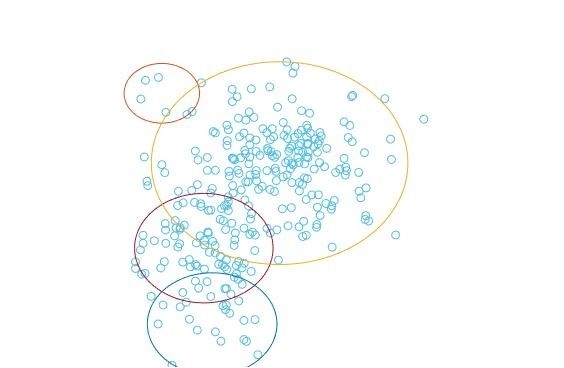}%
}

\caption{Comparison of clustering overlap for a given set of test points with different $d\Sigma_{max}$ values.}

\end{figure}
\par Algorithm 1 shows the pseudocode for the DUCEM scheme. The algorithm takes as an input the drones maximum power $P_{max}$, maximum number of users per cluster $U_{max}$, and the locations of users $\boldsymbol X$. The location and transmitting power of each drone are initialized arbitrarily around the considered area. Furthermore, two thresholds are set: $\varepsilon_1$ which determines the minimum change in drone parameters ($\boldsymbol\mu, \boldsymbol \Sigma$) after which convergence is reached, and  $\varepsilon_2$ which sets the threshold of parameter change below which a new drone would be added if no solution that satisfies power and capacity requirements is yet found. On each iteration the posterior probabilities are obtained and each drone parameters, $\theta$, are updated. Users are then assigned to the drone with lowest resulting $S\!E\!D$ given by (14). The drones 2D locations are then recalculated as the centers of assigned users clusters in step 6. The power of each drone is then obtained by satisfying the $S\!N\!R$ requirement. As long as no solution satisfying the constraints is found the algorithm reiterates. If the maximum change per parameters (step 29) is below $\varepsilon_2$, a new drone is added. The algorithm stops if the change in parameters is below $\varepsilon_1$ and a solution is found. It then returns the settings that result in the highest energy efficiency score (steps 23 - 27).

% \begin{figure}[htp]
%   \centering
%   \subfloat{
%     \includegraphics[clip,width=\columnwidth]{dsigmaNone.jpg}
%     \caption{Overlap for $d\Sigma_{max}=\infty$}}
%     \subfloat{
%     \includegraphics[clip,width=\columnwidth]{dsgima.jpg}
%     \caption{Overlap for $d\Sigma_{max}=0.1$}}
%  \caption{\textcolor{red}{Comparison of clustering overlap for a given set of test points with different $d\Sigma_{max}$ values.}}
% \end{figure}

\begin{algorithm}[H]
 \caption{Drone Users Clustering Expectation Maximization}
 \begin{algorithmic}[1]
 \renewcommand{\algorithmicrequire}{\textbf{Input:}}
 \renewcommand{\algorithmicensure}{\textbf{Output:}}
 \REQUIRE $\boldsymbol X$, $U_{\max}$, $P_{\max}$
 \ENSURE  $M, \;\theta_{opt} = \{ \boldsymbol P,\boldsymbol \mu, \boldsymbol \Sigma\}$, $P_{T_j} \forall j\in \mathcal{M}$, $\boldsymbol W_{M \times U}$ \\
\textit{Initialisation} : Initialize $\theta = \{\boldsymbol P^{(0)},\boldsymbol \mu^{(0)}, \boldsymbol \Sigma^{(0)} \},i=0,$ $maxEE=0$, $\varepsilon_2<\varepsilon_1$
 \\ \textit{LOOP Process}
  \WHILE {($maxd\theta > \varepsilon_1$ or $SolutionFound=0$)}
  \STATE Obtain posterior probabilities $p(j|t)\;\; \forall j\in \mathcal{M}$ 
  \STATE Update $p(j), \boldsymbol \mu_j$ according to (11) and (12) $\forall j\in \mathcal{M}$
  \STATE Obtain $\Sigma_{j_{new}}$ according to (13) $\;\;\forall j\in \mathcal{M}$
  \STATE Allocate users to DSC with minimum SED given in (14) and obtain $\boldsymbol W^i$
  \STATE $\boldsymbol \mu'_j=$ mean ($\boldsymbol W^i_j \boldsymbol X^T$) $\forall j\in\mathcal{M}$
  \STATE calculate $P_{T_j}$ according to $\mathrm{SNR}_T \forall j\in\mathcal{M}$
  \FOR{$j\in \mathcal{M}$}
  \IF {($U_j \;\;\; \leq U_{\max}$ or ${\Sigma_{j_{new}} < \Sigma_j}$)}
  \STATE Update $\Sigma_j = \Sigma_{j_{new}}$
  \ENDIF
  \ENDFOR
  \IF {($maxd\theta < \varepsilon_2$ and $SolutionFound=0 $)}
  \STATE Add new DSC: $M= M+1$
  \STATE Go to step $2$
  \ENDIF
  \STATE $SolutionFound=1$
  \IF {($U_j \;\;\; \geq U_{max}$ or $(P_j>P_{max}))\;\;\forall j\in \mathcal{M} $}
  \STATE $SolutionFound=0$
  \ENDIF
  
  \IF{$SolutionFound=1$}
  \STATE Obtain EE
    \IF{$EE>maxEE$}
    \STATE $maxEE=EE^i$
    \STATE $\theta_{opt}=\theta^i$
    \STATE $\boldsymbol W_{opt}=\boldsymbol W^i$
    \ENDIF
  \ENDIF
  \STATE $maxd\theta=\max\{ \max_{j\in M}\{\mu^i_j-\mu_j^{i-1},\Sigma^i_j-\Sigma_j^{i-1}\}\}$
  \STATE $i=i+1$
  \ENDWHILE
 \RETURN $m, \;\theta_{opt} = \{ \boldsymbol P,\boldsymbol \mu, \boldsymbol \Sigma\}$, $P_{T_j} \forall j\in \mathcal{M}$, $\boldsymbol W_{opt}$
 \end{algorithmic}
 \end{algorithm}

\section{Simulation and Results} \label{Simulation}
\subsection{Users Mobility Model}
For our simulations, we consider that users are moving within groups. This model reflects the scenario in which DSCs deployment is immediately required in mass-events (e.g. concerts, street gatherings, etc.). Other scenarios are also applicable. Each group has a leader moving according to the random waypoint mobility (RWM) model \cite{Bettstetter2003}. Each leader movement is determined by drawing from a memoryless stochastic process with three random variables $D_i,T_{p,i},$ and $V_i$ denoting destination point at instance $i$, pause time at the destination, and traveling speed respectively. 
Furthermore, each leader is surrounded by group members that are moving according to the reference point group mobility model (RPGM) \cite{RPGM}. According to RPGM the movement of group members can be characterized by the respective motion vector:

\begin{equation}
    \boldsymbol V^{(g)}_i(t)=\boldsymbol V^{(g)}_{\mathrm{leader}}(t)+ \boldsymbol {RM}
\end{equation}
where $i, g$ denote the member and group numbers respectively, $\boldsymbol V^g_{leader}(t)$ is the motion vector of group leader $g$ at time instant $t$, and $\boldsymbol {RM}$ is a random vector deviated by group member $i$ from its leader. The movement of group members can be represented by the angle and magnitude deviation between leaders and members' motion vectors. The equations that determine the deviation at time instant $t$ are as follows:

\begin{align}
    |\boldsymbol V_{\mathrm{member}}(t)|&=|\boldsymbol V_{\mathrm{leader}}(t)| + r \;  \phi_v  dV_{max} \\
    \theta_{member}(t)&=\theta_{leader(t)} + r\; \phi_\theta d\theta_{max}
\end{align}
where $r \in [0,1]$ is a random number drawn from a uniform distribution at each group member movement interval, $\phi_v$ and $\phi_\theta$ are the speed deviation ratio and angle deviation ratio respectively, and $dV_{max}$ and $d\theta_{max}$ are the group members maximum speed and angle deviation respectively. Arbitrary settings of $\phi_v$, $\phi_\theta$, $dV_{max},$ and $d\theta_{max}$ yield varied simulation scenarios. 
% \begin{figure}[!t]
% \centering
% \includegraphics[width=1.5in, height=1.5in]{UserMobility.jpg}
% \DeclareGraphicsExtensions.
% \caption{Group member deviation in reference point group mobility (RPGM) model.}
% \label{RPGM}
% \end{figure}

\subsection{Simulation Parameters}
In our simulations we consider an area of $1200 \;\mathrm{m}^2$ for the users' mobility model. The simulation parameters are listed in Table 1. We compare our algorithm results in terms of link reliability and energy efficiency for different values of maximum number of users per cluster $U_{\max}$. The link reliability is given by:
\begin{equation}
    L_{rel}= \Pr\{\Gamma_{u,j} \geq \Gamma_{Th}\}, \;\;\; \forall j\in \mathcal{M},
\end{equation}
where $\Gamma_{u,j}$ is the SINR between user $u$ and drone $j$ given by (4), and $\Gamma_{Th}$ is the SINR threshold above which the link is considered reliable.

\begin{table}[!t]
\renewcommand{\arraystretch}{1.3}
\caption{\textsc{Simulation Parameters}}
\label{table_example}
\centering
\begin{tabular}{|c||c|}
\hline
\textbf{Parameter} & \textbf{Value}\\
\hline
DSCs altitude $h$ & 10 m\\
\hline
Path loss exponent $\lambda$ & 2\\
\hline
DSCs max transmit power $P_{\max}$ & 1 W\\
\hline
Noise power $\sigma_0$ & -100 dBm\\
\hline
Received power at reference distance 1 m $\alpha_0$ & -30 dB\\
\hline
$d\Sigma_{\max}$ & 0.1\\
\hline
Channel bandwidth $B$ & 10 MHz\\
\hline
$\mathrm{SNR}_{T}$ & 12 dB\\
\hline
\end{tabular}
\end{table}

\subsection{K-means Clustering}
K-means clustering is a special case of EM where all components have equal covariances $\Sigma_j$ and each point is assigned to the component with higher posterior probability $p(j|t)$ \cite{MIT}. The algorithm takes in the number of clusters and sample points  $(X,m)$ as an input. For our simulations, we employ K-means clustering with simple modifications to better suit our application. Modifications concern the realistic restriction of the maximum number of users per cluster $U_{\max}$, and maximum transmission power $P_{\max}$.
\par K-means clusters users with the specified restrictions of $U_{\max}$ and $P_{max}$ that DUCEM satisfies. At each iteration, K-means algorithm is repeated 10 times with different initialization determined by Kmeans++ algorithm \cite{kmeans++}. The algorithm is presented in Algorithm 2. As an input, K-means receives the drones maximum power $P_{max}$, maximum number of users per cluster $U_{max}$, and the locations of users $\boldsymbol X$. The algorithm begins with one drone and adds a new one (steps 6- 8) if the constraints are not satisfied. Similarly to DUCEM, the transmission power for each drone is obtained by satisfying the $S\!N\!R$ constraint (step 5) after positioning the drone at the center of the associated users (step 4). Again, the settings that yields the highest $E\!E$ score is returned.

\begin{algorithm}[H]
 \caption{K-means Clustering}
 \begin{algorithmic}[1]
 \renewcommand{\algorithmicrequire}{\textbf{Input:}}
 \renewcommand{\algorithmicensure}{\textbf{Output:}}
 \REQUIRE $\boldsymbol X$, $U_{max}$, $P_{max}$\\
 \ENSURE  $\boldsymbol \mu$, $\boldsymbol W_{opt}$ \\
 \textit{Initialisation}: Initialize $M=1$, $i=0$, $maxEE=0$
 \\ \textit{LOOP Process}
  \WHILE {($i<10$)}
  \STATE Cluster users using K-means with inputs $(\boldsymbol X,M)$
  \FOR{$j\in \mathcal{M}$}
  \STATE $\boldsymbol \mu_j=$ mean ($\boldsymbol W^i_j X^T$)
  \STATE calculate $P_j$ according to $\mathrm{SNR}_T$
  \IF {($ U_j \; > U_{max}$) or ($P_j > P_{\max}$)$ \forall j\in \mathcal{M}$}
  \STATE Cluster users in cluster $j$ using K-means with inputs $(\boldsymbol W^i_j \boldsymbol X^T,2)$
  \STATE Check if sub-cluster $k$ satisfies conditions in $6$. If not, repeat $7$ for cluster $k$
  \ENDIF
  \ENDFOR
  \IF{$EE>maxEE$}
    \STATE $maxEE=EE^{(i)}$
    \STATE $\boldsymbol W_{opt}=\boldsymbol W^i$
  \ENDIF
  \STATE $i=i+1$
  \ENDWHILE
 \RETURN $\boldsymbol \mu_j$, $\boldsymbol W_{opt}$
 \end{algorithmic}
 \end{algorithm}

% \begin{algorithm}[H]
%  \caption{K-means Clustering (non-specified $U_{\max}$)}
%  \begin{algorithmic}[1]
%  \renewcommand{\algorithmicrequire}{\textbf{Input:}}
%  \renewcommand{\algorithmicensure}{\textbf{Output:}}
%  \REQUIRE $\boldsymbol X$, $P_{\max}$ \\
%  \ENSURE  $\boldsymbol\mu$, $\boldsymbol W_{opt}$, $U_{max}$ \\
%  \textit{Initialisation} : Initialize $j=1 ,\;i=0,\; maxEE=0$, $solutionFound=0$
%  \\ \textit{LOOP Process}
%   \WHILE {($i<100$)}
%   \WHILE{$solutionFound=0$}
%   \STATE $solutionFound=1$
%   \STATE Cluster users using K-means with inputs $(\boldsymbol X,j)$
%   \FOR{$m\in \mathcal{M}$}
%   \STATE $\boldsymbol \mu_j=$ mean ($\boldsymbol W^i_j X^T$)
%   \STATE calculate $P_j$ according to $\mathrm{SNR_T}$
%   \IF {($P_j > P_{\max}$)$\; \forall j\in \mathcal{M}$}
%   \STATE Cluster users in cluster $j$ using K-means with inputs $(\boldsymbol W^i_j \boldsymbol X^T,2)$
%   \STATE solutionFound=0
%   \ENDIF
%   \ENDFOR
%   \ENDWHILE
%   \IF{$EE>maxEE$}
%     \STATE $maxEE=EE^{(i)}$
%     \STATE user allocations = current allocations
%     \STATE $U_{\max}= \max(U_{j})$
%   \ENDIF
%   \STATE $i=i+1$
%   \ENDWHILE
%  \RETURN $\boldsymbol \mu_j$, $\boldsymbol W_{opt}$, $U_{\max}$
%  \end{algorithmic}
%  \end{algorithm}

\subsection{Results}
\begin{figure}[htbp]
\centering
\includegraphics[width=2.9in]{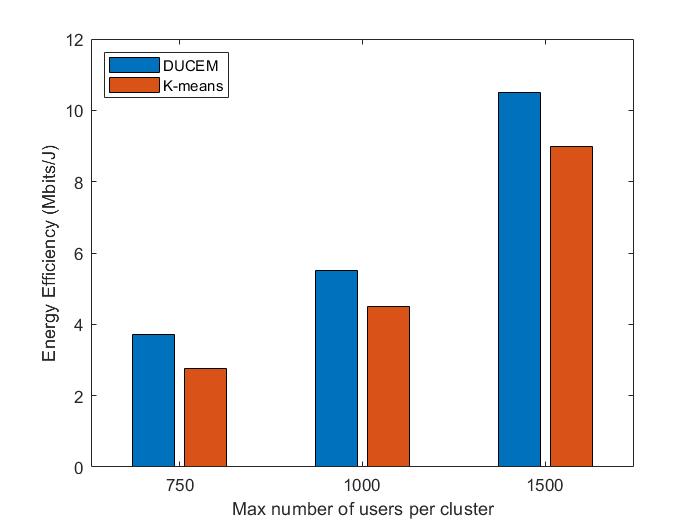}
\DeclareGraphicsExtensions.
\caption{System energy efficiency (EE) against the maximum allowed number of users per cluster ($U_{max}$).}
\label{EE}
\end{figure}

\begin{figure}[htbp]
\centering
\includegraphics[width=2.9in]{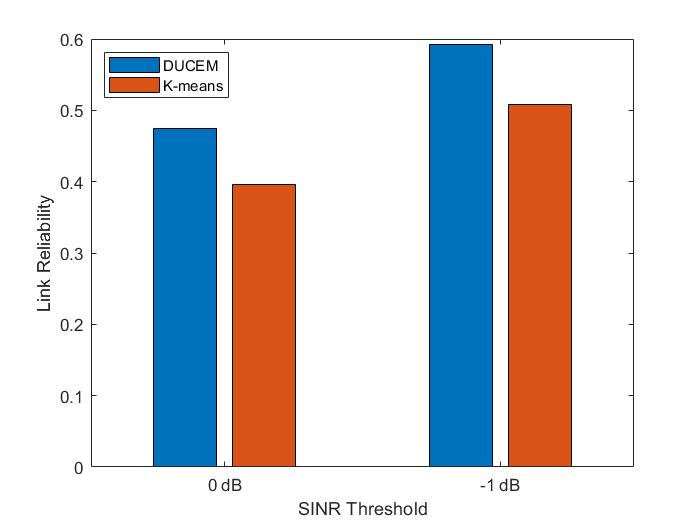}
\DeclareGraphicsExtensions.
\caption{Link reliability ($L_{rel}$) against SINR threshold ($\Gamma_{Th}$).  $U_{max}=1500$.}
\label{Reliability}
\end{figure}

Fig. \ref{EE} shows the obtained EE for different values of $U_{max}$. On average, the EE of the system is improved by $25\%$. Furthermore, the achieved link reliability is obtained for two different values of $\Gamma_{Th}$. The results show that DUCEM provides, on average, $18.3\%$ better link reliability. Fig.  \ref{Reliability} plots $L_{rel}$ results for $U_{max}=1500$. 
\par We observe that DUCEM results are indeed more optimal than the widely used K-means clustering mechanism. The improvement in performance is almost constant for different values of $U_{max}$. This is expected since both algorithms do not directly optimize inter-cell interference which is expected to have greater effect with increasing number of drones (i.e. decreasing value of $U_{max}$). Employing DUCEM with the mechanisms mentioned in Section \ref{Related Work} is expected to speed up convergence and achieve better results which will be the topic of our future work. 
\section{Conclusion}
\label{conclusion}
DSCs provide a suitable solution for providing wireless coverage. However, their deployment problem requires consideration of various aspects such as co-channel interference, 3D placement, user clustering, and transmit power minimization. The problem of user clustering for DSCs has received limited attention. In this paper, a user clustering algorithm was proposed that takes into account the aforementioned aspects of the problem. The numerical results verified that the proposed algorithm enhances system performance in terms of energy efficiency and link reliability. The results of DUCEM can provide the initial clustering upon which further performance improving mechanisms can be implemented. To further improve the performance of DUCEM, the problem of inter-cell interference should be tackled and optimized. Other aspects to be considered and optimized for are users handover and the drone's propuslion energy.

% conference papers do not normally have an appendix

% use section* for acknowledgment

% trigger a \newpage just before the given reference
% number - used to balance the columns on the last page
% adjust value as needed - may need to be readjusted if
% the document is modified later
%\IEEEtriggeratref{8}
% The "triggered" command can be changed if desired:
%\IEEEtriggercmd{\enlargethispage{-5in}}

% references section

% can use a bibliography generated by BibTeX as a .bbl file
% BibTeX documentation can be easily obtained at:
% http://mirror.ctan.org/biblio/bibtex/contrib/doc/
% The IEEEtran BibTeX style support page is at:
% http://www.michaelshell.org/tex/ieeetran/bibtex/
\bibliographystyle{IEEEtran}
% argument is your BibTeX string definitions and bibliography database(s)
\bibliography{references.bib}
%
% <OR> manually copy in the resultant .bbl file
% set second argument of \begin to the number of references
% (used to reserve space for the reference number labels box)

% \begin{thebibliography}{1}

% \bibitem{IEEEhowto:kopka}
% H.~Kopka and P.~W. Daly, \emph{A Guide to \LaTeX}, 3rd~ed.\hskip 1em plus
%   0.5em minus 0.4em\relax Harlow, England: Addison-Wesley, 1999.

% \end{thebibliography}

% that's all folks
\end{document}